# Preprint

# The Weak Convergence of TCP Bandwidth Sharing

Wolfram Lautenschlaeger

Nokia Bell Labs
Stuttgart, Germany

**Abstract.** TCP is the dominating transmission protocol in the Internet since decades. It proved its flexibility to adapt to unknown and changing network conditions. A distinguished TCP feature is the comparably fair resource sharing. Unfortunately, this abstract fairness is frequently misinterpreted as convergence towards equal sharing rates. In this paper we show in theory as well as in experiment that TCP rate convergence does not exist. Instead, the individual TCP flow rate is persistently fluctuating over a range close to one order of magnitude. The fluctuations are not short term but correlated over long intervals, so that the carried data volume converges rather slowly. The weak convergence does not negate fairness in general. Nevertheless, a particular transmission operation could deviate considerably.

## 1 Introduction

The Transmission Control Protocol (TCP) is used for reliable data transmission over packet switched networks. The TCP transmitter splits the data into segments, encapsulates them into IP packets, and sends them to the receiver. The receiver reassembles the data from the incoming segments. Lost packets are detected by means of sequence numbers. The receiver signals back to the transmitter the successful reception of data by acknowledgement packets (ACK). Duplicate and selective acknowledgements (SACK) are used to signal packet loss. The transmitter in turn retransmits the previously lost packets. Packet transmission and the acknowledgement back take some time, in particular for forwarding, propagation, queuing, and processing in both directions, which is altogether called the Round Trip Time (*RTT*).

TCP restricts its own transmission rate for congestion control. This is done by a congestion window (*cwnd*) that at any time limits the amount of data that has been sent out, but that has not been acknowledged yet (the so called data in flight). This way the transmission rate is limited to *cwnd* divided by *RTT* (i.e. packets/s). Since the transmitter typically does not know the available transmission capacity along the path, it continuously probes for more bandwidth by gradually increasing the *cwnd*. In contrast, as soon as packet loss is signaling congestion, the *cwnd* is shrunk, typically by half. The succession of slow increases and abrupt decreases (sawtooth oscillation) eventually stabilizes the transmission rate at the limit of the available transmission capacity [1].

If several TCP flows share the same limited transmission resource, then each of them tries to get more of the shared resource at the cost of the others. Under the assumption of similar conditions, it is natural to expect convergence of flow rates, eventually leading to equal sharing. A first proof of rate convergence was given in [2]. The convergence speed was analyzed in [3], yielding a 98% convergence towards fair sharing rate within seven sawtooth cycles. The convergence time into an ε-environment of the fair sharing rate was frequently used for characterization of different TCP flavors [4], [11].

Unfortunately, and in opposite to what the mentioned papers suggest, something like a monotonic TCP rate convergence towards the fair sharing rate does not exist. In this paper we show that the rate of a TCP flow walks randomly around its fair sharing rate. It deviates down to 1/3 and up to the 3 fold of that rate, altogether within a 1:10 span of possible flow rates. The rate variations are not short term, so that no significant averaging can be observed up to the minutes range, and it takes hours to get stable average values. Why the theories on TCP rate convergence missed that effect? The problem is typically linked to a premature average assumption in the course of modelling the bandwidth sharing process, which finally proves only convergence of an expectation value of the flow rate. However, the expectation value tells little about the actual rate, its distribution, and its realization over time. What remains undisputed with this paper is the equal *cumulative* rate sharing over infinite time, in contrast to other potential assumptions like e.g. "winner takes all".

The paper is structured as follows: After the introduction we elaborate in section 2 the theoretical TCP flow rate distribution at random packet loss. In section 3 we reproduce the distribution in an experiment with real network equipment. Then we show that bandwidth sharing creates quite similar distributions like at purely random loss. Furthermore we investigate the temporal aspects and show that rate deviations are not short term, but much larger than the round trip time. In section 4 we illustrate the consequences of the weak convergence for streaming applications and for the flow completion times of typical short lived flows. We further discuss the implications for Active Queue Management (AQM) and the related experimental work. Section 5 summarizes the findings.

## 2    TCP Bandwidth Theory

### 2.1    Basic TCP equations

TCP operation in congestion avoidance mode as explained in the introduction follows a number of well-known formulas that we recall here for reference:

With the maximum segment size *MSS* (roughly the packet size) in bits and the round trip time *RTT*, the bit rate *b* of a congestion window *cwnd* limited TCP flow is

$$b = \frac{MSS \cdot cwnd}{RTT} \quad (1)$$

For TCP Reno [18] the gradual additive increase of *cwnd* during congestion avoidance per *RTT* is

$$cwnd \leftarrow cwnd + 1 \quad (2)$$

In reality it is *cwnd* ← *cwnd* + 1/*cwnd* per received acknowledgement. Since *cwnd* segments are in flight, *cwnd* acknowledgements return during one RTT, which yields Eq. 2. We will see later that the real increase is slower due to the delayed acknowledgments. Other TCP flavors like Cubic have variable and partially larger growth rates.

The abrupt multiplicative *cwnd* reduction due to loss detected follows

$$cwnd \leftarrow \frac{cwnd}{2} \quad (3)$$

Here also variations are possible, e.g. Cubic does a smaller reduction according to *cwnd* ← 0.7·*cwnd*.

The steady state performance of a TCP flow at certain packet loss probability $P_{loss}$ has been multiply derived [5], [6], [7]. Taking into account the delayed acknowledgment ratio $a = 2$ we get for the expected *cwnd*:

$$\mathrm{E}[cwnd] = \sqrt{\frac{3}{2a}} \frac{1}{\sqrt{P_{loss}}} \quad (4)$$

Together with Eq. 1 the expected flow bit rate *b* is

$$\mathrm{E}[b] = \sqrt{\frac{3}{2a}} \frac{MSS}{\sqrt{P_{loss}} RTT} \quad (5)$$

Equation 5 can be reverted: Bandwidth sharing with certain flow bit rate *b* must result in a corresponding packet loss ratio $P_{loss}$.

The behavior of TCP Cubic is slightly different. We recall here the formula from the original Cubic paper [11]:

$$\mathrm{E}[cwnd_{cubic}] = 1.17 \cdot \left(\frac{RTT}{P_{loss}}\right)^{\frac{3}{4}} \quad (6)$$

where *RTT* is given in seconds.

## 2.2 Origin of Packet Loss

Packets are almost exclusively lost due to buffer overflow in intermediate nodes. Other sources of packet loss like bit errors or link degradation are out of scope of TCP for different reasons: Wireline links operate at bit error rates below $10^{-12}$, thus causing CRC errors on packet level by orders of magnitude below typical TCP loss rates. Wireless links use link layer handshake protocols for packet delivery to hide the drastic loss rates from higher layers. TCP sees only throughput and delay degradations that in turn might induce buffer overflow and retransmission time outs, but no packet drops.

Buffer overflow occurs due to deterministic queue filling by TCP sources, due to stochastic reasons (typically modelled by M/D/1 queues or some kind of burstiness), or, in practice, due to a combination of both. In the simplest case, one TCP flow crossing one bottleneck link, the process is fully deterministic: If the link is already loaded at 100%, any further *cwnd* increase grows the queue before the link until it overflows the available buffer space. Finally, at overflow, one packet is dropped, TCP reduces its *cwnd* by half, and the queue size goes down, accordingly. It looks like the *cwnd* is oscillating between a maximum and half that value. Simple TCP theories are built on that assumption. Nevertheless, it is not the *cwnd* maximum, but the queue size that triggers the loss. It is just that both go synchronized in the single flow case.

If two (or more) TCP flows cross the same bottleneck, the initial picture looks similar: The cumulative increase of *cwnd* in both sources grows the queue. But then, at overflow, one or two packets are dropped. It is not assured that both flows catch a loss. First of all it could be only one drop. Second, if two packets are dropped, they could belong to one and the same flow, leaving the other one untouched. For the queue it does not matter. It is sufficient that one source reduces its *cwnd* to get away from the buffer limit. In either way, it is not the rule that both flows reduce their *cwnd* at the same time. The two flows, even if started synchronous, move apart from each other. One continues to grow its *cwnd*, while the other one resumes its *cwnd* growth at only half that level. That inequality is going to be resolved at next drop cycle, right? Unfortunately not. The *cwnd* size does not matter for the drop; only the queue matters, which is identical for both flows. Admittedly, the flow with the larger *cwnd* sends more packets than the other flow. This increases its probability to catch a drop, if one occurs. In the long run this results in the weak convergence. But at the moment it is not unlikely that the flow with the smaller *cwnd* catches once more the drop, and shrinks its *cwnd* further, while the larger flow continues to grow.

A detailed mathematical analysis of the bandwidth sharing process can be found in [7]. As one of the results, with a tail drop queue, approximately half of the competing flows are affected by a single buffer overflow event. For this paper it does not really matter how many packets are dropped at once and why. The only required plausible insight is that, once drops occur, not all but only a random subset of flows is affected. This is the main difference to the misleading convergence analysis of [2] and [3].

## 2.3 Flow Rate at Random Packet Loss

In this section we investigate the probability distribution of TCP flow rates at random drop, irrespective of a particular bandwidth sharing assumption. We presume that every packet of a TCP flow is dropped at probability $P_{loss}$ with no regard of preceding losses, which results in a Poisson loss process. In context of bandwidth sharing the assumption of a Poisson loss process *per flow* is not arbitrary. A proof in [14] (section 7.7.1) indicates that for increasing flow numbers the loss process *per flow* converges towards independence of losses, no matter what loss distribution holds for the whole aggregate.

We analyze TCP Reno with Delayed Acknowledgements [15] but without Appropriate Byte Counting (ABC) [16]. Delayed ACK means the receiver sends less than one ACK per received segment for efficiency reasons, typically one ACK per two segments. ABC was intended to compensate the delayed ACK effect on the *cwnd* handling. However, in the Linux kernel the ABC feature was switched off by default since years and recently it has been removed completely [17]. We account for the uncompensated effect of delayed ACK by the acknowledgement ratio a = 2 (segments per ACK).

The expected flow bit rate is given by Eq. 5. The probability distribution of the flow bit rate can be obtained by investigating the evolution of the congestion window *cwnd* as a continuous Markov chain. (We stick here to a method from [8].) Figure 1 shows a fragment of the Markov chain, where the state nodes correspond to the actual *cwnd* size, and transition arcs correspond to conditional transition rates between the states. An arrow from node *i* to node *j*, labeled by rate $r_{ij}$, indicates that, if *cwnd* is in state *i*, this state is left towards state *j* at rate $r_{ij}$. The absolute transition rate depends on the probability $p_i$ to find *cwnd* in state *i*. Thus, the absolute rate from *i* to *j* is $p_i \cdot r_{ij}$. If we assume for a moment that in a given state the sum of arriving rates is larger than the sum of departing rates, obviously its probability would go up. Since probabilities are static by definition, we need to find the equilibrium, where for all nodes the sum of arriving rates equals the sum of departing rates. The equilibrium can be calculated as follows:

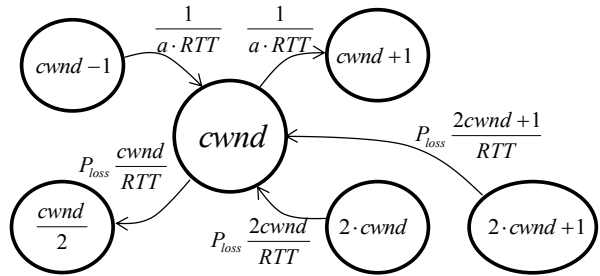

**Figure 1**. Fragment of the congestion window state diagram

For the upper part of Figure 1 holds: The *cwnd* is incremented by an amount of 1/*cwnd* for every arriving ACK. Since *cwnd* packets are in flight, after one *RTT* the total *cwnd* increment should be one per *RTT*. Due to the uncompensated delayed ACKs, however, only 1/*a* (i.e. half) of the 1/*cwnd* increments are executed. Hence, the rate of *cwnd* increments is 1/*a* per one *RTT*; the transition rate from *cwnd* to *cwnd*+1 is:

$$r_{cwnd \to cwnd+1} = \frac{1}{a \cdot RTT} \qquad (7)$$

For the lower part of Figure 1 holds: The actual packet rate is $r_{pack}=cwnd / RTT$. Packets are lost at probability $P_{loss}$. Correspondingly the packet loss *rate* (lost packets per second) is $r_{loss}=P_{loss} \cdot r_{pack}$. Thus the *cwnd* halving rate (transition rate from state *cwnd* to state *cwnd*/2) is:

$$r_{cwnd \to \frac{cwnd}{2}} = P_{loss} \frac{cwnd}{RTT} \tag{8}$$

In fact, this reflects that, even though the drop probability $P_{loss}$ is equal for all flows, the hit *rate* of a particular flow depends on the amount of packets sent, so that larger flows are more likely affected than smaller ones.

The equilibrium equation of state *i*, where incoming and outgoing rates are equal, is

$$(1/a)p_{i-1} + 2iP_{loss} \cdot p_{2i} + (2i+1)P_{loss} \cdot p_{2i+1} = (P_{loss}i + 1/a)p_i \tag{9}$$

The state probabilities $p_i$ of *cwnd* to be in state $i \in [1, cwnd_{max}]$ form a set of linear equations. In matrix notation the corresponding state probability vector $P_{cwnd} = (p_1, p_2, \cdots, p_{cwnd_{max}})^T$ fulfills following equilibrium equation:

$$P_{cwnd} = A \cdot P_{cwnd} \tag{10}$$

The extreme cases need special care: TCP limits *cwnd* to at least 2 since otherwise the loss detection by duplicate ACKs would not work anymore. As consequence state 2 can be left only by increment, but not by rate halving. Furthermore state 2 can additionally be reached from state 3 by halving. At the other end, the maximum *cwnd* can be left only by halving, but not by increment.

With the shortcut $P=a \cdot P_{loss}$ the transition matrix $A$ (with e.g. $cwnd_{max}=9$) looks as follows:

$$A = \begin{bmatrix} 0 & 0 & 0 & 0 & 0 & 0 & 0 & 0 & 0 \\ 1 & 0 & \frac{3P}{1} & \frac{4P}{1} & \frac{5P}{1} & 0 & 0 & 0 & 0 \\ 0 & \frac{1}{1+3P} & 0 & 0 & 0 & \frac{6P}{1+3P} & \frac{7P}{1+3P} & 0 & 0 \\ 0 & 0 & \frac{1}{1+4P} & 0 & 0 & 0 & 0 & \frac{8P}{1+4P} & \frac{9P}{1+4P} \\ 0 & 0 & 0 & \frac{1}{1+5P} & 0 & 0 & 0 & 0 & 0 \\ 0 & 0 & 0 & 0 & \frac{1}{1+6P} & 0 & 0 & 0 & 0 \\ 0 & 0 & 0 & 0 & 0 & \frac{1}{1+7P} & 0 & 0 & 0 \\ 0 & 0 & 0 & 0 & 0 & 0 & \frac{1}{1+8P} & 0 & 0 \\ 0 & 0 & 0 & 0 & 0 & 0 & 0 & \frac{1}{9P} & 0 \end{bmatrix} \tag{11}$$

Since Eq. 10 is a homogeneous system, we replace for a numeric solution one of the component equations by the normalizing condition $\Sigma\ p_i =1$. Then, the bit rate distribution is the *cwnd* state probability vector, scaled according to the TCP throughput Equation 1.

In Figure 2, the graph labeled "theory" shows the numerically evaluated bit rate probability density of a TCP flow. A similar result has been published already in [9].

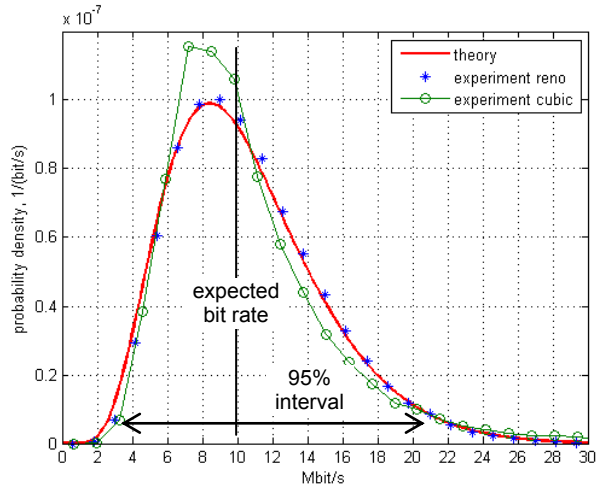

**Figure 2** Bit rate distribution of a TCP flow at random packet loss

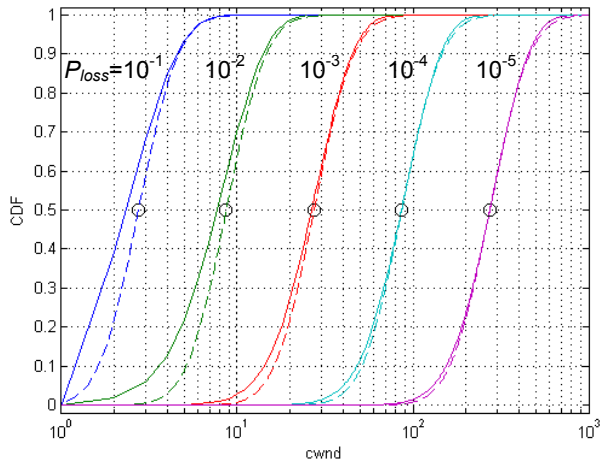

**Figure 3** . Numerically calculated CDF of the congestion window *cwnd*; dashed lines are the log normal CDF of Eq. 12; markers show the *cwnd* expectation value of Eq. 4

The flow bit rate distribution has a substantial spreading. The 95% interval is ranging roughly from less than 40% up to more than double the expected rate. Since we are calculating equilibrium probabilities, this distribution holds over an infinite span of time. There is no room for further convergence towards the expected rate. The spreading statement is quite strong. It holds for a wide range of loss probabilities. Figure 3 shows numerically calculated cumulative distribution functions (CDF) of the congestion window. The relative spreading is fairly constant over 5 decades of $P_{loss}$. For better understanding we complement the graphs with plots of the log normal distribution

$$F(cwnd) = \Phi\left(\frac{\ln\frac{cwnd}{E[cwnd]}}{\sigma}\right), \quad (12)$$

where $\Phi$ is the cumulative standard normal distribution function, $E[cwnd]$ – the expectation value of *cwnd* according to Eq. 4, and $\sigma = 0.41$ – the constant logarithmic standard deviation.

Obviously the *cwnd* and derived thereof the TCP flow bit rate have a stable spread around the expectation value. The relative spread is nearly invariant of the packet drop probability; it reaches an order of magnitude; and it does not vanish over time.

## 3 Experimental Evaluation

In this section we verify, if the theoretically calculated bit rate distribution can be observed in practice. We present an experiment with just one TCP flow in an uncongested network, but with artificial random packet drop, thus reproducing the scenario of the theoretical analysis. Then we compare the results with bandwidth sharing experiments with 2, 3, and 10 concurrent flows, but without artificial packet drop. Here we show that the bit rate spreading is comparable with the random drop case. Finally we investigate how long flow rate deviations persist and how fast deviating flow rates return towards their fair sharing value.

The experiments have been executed on a networking testbed of Linux servers and Ethernet switches. All connections are 10G Ethernet with all TCP offloading features disabled. TCP parameters, if not specially mentioned, are the defaults of Linux kernel 3.16. The conditions are chosen such that each flow has a bit rate expectation value of $E[b]=10$Mbit/s. This way we exclude bit rate dependent transmitter or receiver specific variations from our experiments. Round trip time, if not stated otherwise, was $RTT = 100$ms. Duration of each run was 12 hours. The total throughput of all bandwidth sharing experiments was above 99%.

### 3.1 Random packet loss

In this experiment we use a single TCP flow. The transmitted packets are randomly dropped by a specially adapted `iptables` rule. The rule draws for every arriving packet a uniformly distributed random number between 0 and 1. The packet is dropped if the random number is smaller than the requested drop probability. The 10G Ethernet network is loaded in average at 10Mbit/s so that no queuing or congestion impact is to be expected. We performed the experiments with TCP Reno (the reference) and TCP Cubic as the current Linux default. To reach the 10Mbit/s target we used a drop probability according to Eq. 5 for TCP Reno, and for TCP Cubic according to Eq. 6 (i.e. $P_{reno}=1.1\cdot10^{-4}$, $P_{cubic}=3.4\cdot10^{-4}$).The flow rate distribution is captured by counting the carried bytes in one second intervals. The count values are than accumulated in the bins of a histogram. More than 43,000 count values per experiment (12 hours) have been obtained to get a stable estimation of the distribution function.

Figure 2 of section 2.3 shows besides the theoretical distribution a comparison with the experimental results. Obviously the TCP Reno experiment reproduces exactly the theoretically calculated flow rate distribution. Remaining deviations are so small that they easily can be attributed to the finite duration of the experiment. The experiment with TCP Cubic shows a small deviation. Nevertheless, the spreading of the distribution is similar to TCP Reno.

### 3.2 Bandwidth sharing

In this experiment we used 2, 3, or 10 identical TCP flows that share a common bottleneck of 20, 30, or 100 Mbit/s, respectively, which results always in the same target rate of 10Mbit/s per flow. The bottleneck and the corresponding queue are created by the traffic control subsystem of an intermediate Linux server (the `tc qdisc` command). The buffer size for the bottleneck queue was chosen according to the bandwidth delay product rule (BDP). Figure 4 shows the flow rate distribution of the bandwidth sharing experiments, again in comparison to the theoretical distribution at random drop. The bit rate distribution has been measured for one arbitrarily picked flow out of the 2, 3, or 10 flows by the same histogram method as in section 3.1.

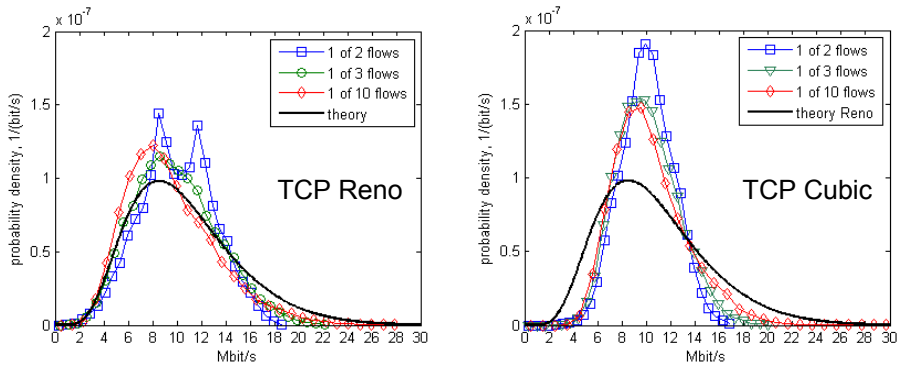

**Figure 4** Experimental distribution of bandwidth sharing TCP flow rates

The shape and spread of the curves is similar to the theoretical distribution. TCP Cubic shows a slightly more concentrated distribution around the expected bit rate of 10Mbit/s. Nevertheless, in all cases the spread of flow rates is so large that deviations below half of the expectation value and above double that value are possible. Even after 12 hours of continuous bandwidth sharing there is no sign of rate convergence.

Table 1 summarizes the experimental flow rate distributions by their mean and the 5%, 50%, and 95% quantiles.

**Table 1.** Flow rate statistics

|       |                          | quantiles, Mbit/s |      |      | mean, Mbit/s |
|-------|--------------------------|-------------------|------|------|--------------|
|       |                          | 5%                | 50%  | 95%  |              |
| Reno  | random drop (numeric)    | 4.7               | 10.0 | 19.0 | 10.7         |
|       | random drop (experiment) | 4.9               | 10.0 | 18.7 | 10.7         |
|       | 1 of 2 flows             | 5.0               | 10.0 | 15.0 | 10.0         |
|       | 1 of 3 flows             | 4.7               | 9.6  | 16.0 | 9.9          |
|       | 1 of 10 flows            | 4.5               | 8.9  | 16.6 | 9.5          |
| Cubic | random drop (experiment) | 5.0               | 9.4  | 20.0 | 10.6         |
|       | 1 of 2 flows             | 6.5               | 10.0 | 13.6 | 10.0         |
|       | 1 of 3 flows             | 6.3               | 9.8  | 14.6 | 10.0         |
|       | 1 of 10 flows            | 6.1               | 9.8  | 16.0 | 10.3         |

### 3.3 Duration of rate variations

A frequently raised argument for a technical convergence is that the TCP flow rate might be highly unsteady or even bursty at time scales of one *RTT* or below, but that these variations quickly vanish if looking at the duration of typical TCP flows of few *RTT*s. The argument silently assumes that there is no correlation over a distance of more than a few *RTT*s. In this section we investigate how fast the average rate over certain interval duration converges towards the expectation rate.

We repeated all experiments of the previous sections but with different interval settings, i.e. we counted the carried bytes not only in intervals of 1 second but additionally in intervals of 4, 16, 30, 60, 120, 300, and 600 seconds over a total time of 12 hours. From the series of count values we calculated the standard deviation of the flow rate at the particular interval settings. Figure 5 shows the results. It reproduces the impression of the previous sections that the flow rate variations slightly grow with the number of flows, but still stay below the value at purely random loss, and that they are larger in general for TCP Reno than for TCP Cubic. As expected, the standard deviation shrinks with increasing interval duration. However, the decline is very slow. It remains negligible up to 20 – 30 second intervals, and even for 10 minute intervals the standard deviation stays in the range of 10% of the mean (10Mbit/s).

The graphs also justify our experimental approach for verification of the theory. In fact, the theory of section 2 is correct in a strong sense for intervals of one round trip, including the queuing delay, i.e. variable 100 – 200ms, depending on the actual queue size. In contrast, the experimental data have been obtained as data volume carried over constant intervals of one second. In our case the graphs are comparably flat in the neighborhood of one second, so that the interval mismatch with the theory can be accepted.

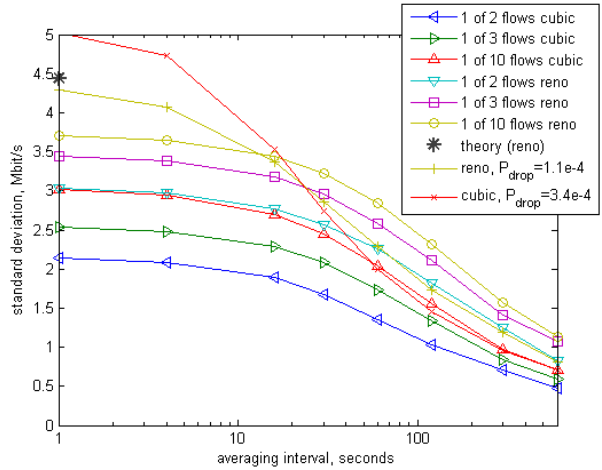

**Figure 5**  Standard deviation of short term average rates at different interval durations; bandwidth sharing and random drop experiments

In a further experiment we investigated the impact of the round trip time. Instead of *RTT* = 100ms (the default *RTT* in this study), we used an *RTT* of only 10ms and a corresponding bandwidth delay product (BDP) sized buffer. The results are shown in Figure 6.

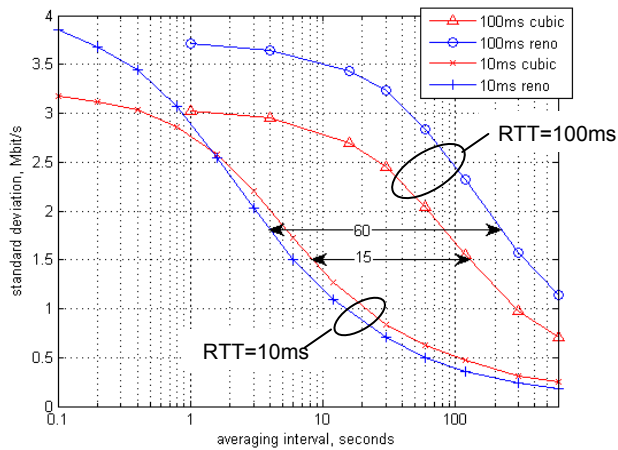

**Figure 6**  Impact of the RTT on the convergence

As expected, the convergence slope shifts left, towards smaller intervals. The shift is much more pronounced for TCP Reno than for Cubic, so that the mutual order reverts. The shift for Reno is by a factor of 60, which can be weakly associated with the theoretical sawtooth interval that scales quadratic with the *RTT*, i.e. a shift of 100 could be expected. The shift for Cubic is much smaller, by a factor of 15, which is in line with Cubic's original intention to make TCP less *RTT* sensitive. Nevertheless, the reduction is even larger than what Cubic's performance Equation 6 might suggest. We verified that by measuring the actual packet loss rates and comparing them with the theory. The values fit well for all experiments, except the 10ms Cubic case. Here Cubic drops 5 times more packets than required according to Eq. 6. The reason for this mismatch is a fallback heuristic in the Cubic algorithm (a bit misleadingly named `tcp_friendliness`): According to the original Cubic paper [11] it approximates, in addition to its own *cwnd*, the corresponding TCP Reno window and takes the larger of the two windows.

**Table 2.** RTT dependence of convergence

|       | RTT   | $P_{loss}$ | | sawtooth interval | 50% convergence interval | ratio |
|-------|-------|--------|------------|-------------------|--------------------------|-------|
|       |       | theory | experiment |                   |                          |       |
| reno  | 10ms  | 3.8e-3 | 3.3e-3     | 0.37s             | 4s                       | 11    |
|       | 100ms | 3.8e-5 | 4.0e-5     | 29.5s             | 220s                     | 7.5   |
| cubic | 10ms  | 6.2e-4 | 2.8e-3     | 0.42s             | 9.5s                     | 22    |
|       | 100ms | 2.9e-4 | 2.5e-4     | 4.7s              | 130s                     | 27    |

The experimental results are summarized in Tab. 2. The sawtooth interval is calculated from the experimental loss ratio. The 50% convergence interval is the duration where the carried data volume fluctuates just half as much as at the smallest intervals. The last column is the ratio between convergence interval and sawtooth interval.

## 4   Consequences

The bit rate of a bandwidth sharing TCP flow does not converge at all. Instead it walks randomly around its fair sharing expectation value. Deviations are not small; they go down to less than half of the fair sharing rate, and up to more than double that value. Deviations are not short term; they last thousands of round trip times; in our experiments many minutes. And the deviations do not attenuate over time; their spread stays the same after many hours of continuous bandwidth sharing. Figure 7 illustrates these facts for the last 10 minutes of a 12 hours bandwidth sharing experiment with just two flows. (The link was loaded all the time at constant 20Mbit/s; the two flows complemented each other at any time.)

The effect is relevant for streaming applications, like video streaming. These applications rely on a continuous arrival of new content. They need sufficient margins to cope with the rate variations or flatten the arrival by a playout buffer. Figure 5 gives an impression of how long a playout buffer needs to store to get a reasonable flattening effect.

The effect is also relevant for the flow completion time of finite TCP flows. In general it is assumed that a new flow entering a congested link with *N*-1 pre-established flows grabs a 1/*N* fraction of the link bandwidth and completes accordingly. However, the actual flow rate variates according to Figure 4. If the variations persist longer than the flow duration, the actual flow completion time gets a similar spread, i.e. ranging from half the expected duration up to more than double that time.

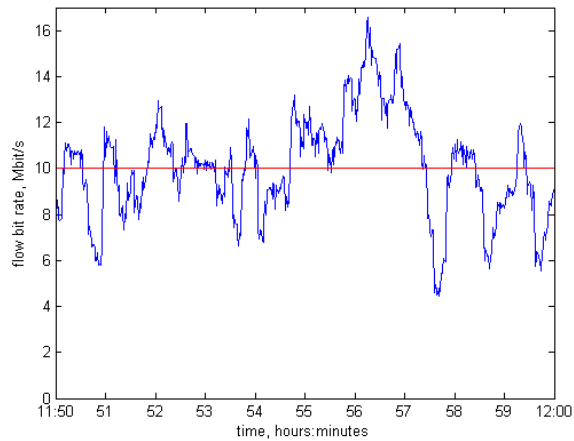

**Figure 7** Random walk: Last minutes after 12 hours of continuous bandwidth sharing; one of two TCP Cubic flows at RTT=100ms in 20Mbit/s link bandwidth

In the experiment of Figure 8 we run 9 long lived TCP flows over a link of 100Mbit/s. Then we launched repeatedly a 10$^{th}$ short lived flow with a data volume of 12Mbyte. The expected rate is 10Mbit/s, the expected duration 10 seconds. The displayed four shots carry all the same data volume, but it takes between 7 and up to 17 seconds till completion. In a more exhaustive experiment with 2500 repetitions, 5% of the flows take less than 8 seconds, whereas another 5% take more than 22 seconds till completion.

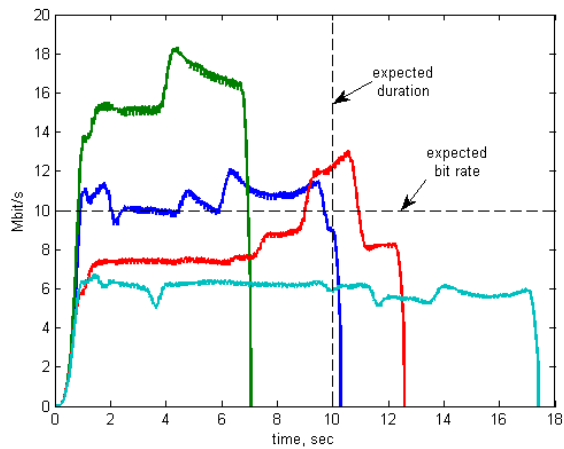

**Figure 8** Transmission of 12 Mbyte at expected fair sharing rate of 10Mbit/s; 4 independent shots in an otherwise identical set-up

The weak convergence bears more implications on TCP rate control. It seems to be impossible to directly control a TCP flow rate by applying random packet drop according to the well-known TCP bandwidth formula Eq. 5. The reaction is too fuzzy, and if relying on a cumulative effect, the response is much too slow. Existing Active Queue Management (AQM) solutions like Random Early Detection (RED) [12] always incorporate a queue. That queue is not acting just as an averaging device. Instead, in the first instance it establishes equilibrium between the congestion windows of all involved transmitters and the queueing delay, this way stabilizing the total rate. Only secondarily RED confines the equilibrium queue to the available buffer space by random dropping. Since the queue is unique for all flows, this approach stabilizes only the total rate of all flows. The particular flow continues to spread out as of Figure 4.

Since the weak convergence is rooted in the arbitrary assignment of packet drops to the affected flows, it is unlikely to find AQM mitigation without some kind of *flow notion*. In normal packet nodes this is not the case, impractical, or at least undesirable due to the noticeable additional effort. For further reading we refer to the well-known queueing disciplines WRR or SFQ [13] and recent derivates by Linux kernel modules `sch_fq`, `sch_fq_codel`.

Special care is required in measurement experiments for characterization of novel TCP and queuing approaches. Metrics like the $\varepsilon$-convergence time of [4] are inherently undefined, since a flow that reached the $\varepsilon$ environment of the expected rate is not guaranteed, not even likely, to stay in that $\varepsilon$ environment. Experiments that claim such convergence anyway likely stopped prematurely at the first visit. In general, the experimental acquisition of per flow metrics requires extremely long observation times of hours or days, rather than seconds or minutes. Nonetheless, this must not be confused with global metrics, characterizing the combined effect of all involved flows like total rate, queue size, or drop ratio. These metrics usually converge much faster.

## 5    Summary

TCP is able to fill a network bottleneck at 100% of its transmission capacity. If multiple flows share the same bottleneck, then the available bandwidth is distributed between the flows in a comparably fair way: (1) None of the flows is able to monopolize the available bandwidth. (2) None of the flows starves. Under uniform conditions (same RTT, same TCP flavor) the rate expectation and the long term average are equal for all sharing flows. The carried data volume of the flows converges to equal values at infinity.

In this paper we investigate to which extent this "equal sharing" proposition can be applied to technically relevant conditions. We show that the actual rate of a particular flow does not converge at all. It deviates randomly down to one third and up to three fold of its expected rate. The random deviations do not attenuate over time, neither in theory nor in experiment. In our experiments they appear even after many hours of continuous bandwidth sharing. And the deviations are long lasting. Their correlation span is many times larger than the Round Trip Time or the TCP sawtooth interval. Accordingly, the carried data volume converges only slowly after thousands of RTT. The findings have been theoretically derived and subsequently verified by comprehensive series of bandwidth sharing experiments in a test bed of Ethernet servers and switches.

**Acknowledgement**

This work has been funded in part by the German Bundesministerium für Bildung und Forschung (Federal Ministry of Education and Research) in scope of project SASER under grant No. 16BP12200.